\documentclass[12pt]{iopart}

\usepackage{iopams}
\usepackage{setstack}

\usepackage[english,british]{babel}
\usepackage{graphicx}
\usepackage{hyperref}
\usepackage{epstopdf}
\usepackage{color}
\usepackage[utf8]{inputenc}
\newcommand{\eg}{e.\,g.\ }

\newcommand{\cf}{cf.\ }

\begin{document}
\title[Self-ordering in broadband radiation]
{Tailored long range forces on polarizable particles by collective scattering of broadband radiation}
\author{D. Holzmann and H. Ritsch}
\address{Institute for Theoretical Physics, University of Innsbruck, Technikerstra\ss e 25, A-6020 Innsbruck, Austria}
\ead{Daniela.Holzmann@uibk.ac.at}
\begin{abstract}

Collective coherent light scattering by polarizable particles creates surprisingly strong, long range inter-particle forces originating from interference of the light scattered by different particles. While for monochromatic laser beams this interaction decays with the inverse distance, we show here that in general the effective interaction range and geometry can be controlled by the illumination bandwidth and geometry. As generic example we study the modifications inter-particle forces within a 1D chain of atoms trapped in the field of a confined optical nanofiber mode. For two particles we find short range attraction as well as optical binding at multiple distances. The range of stable distances shrinks with increasing light bandwidth and for a very large bandwidth field as e.g. blackbody radiation we find a strongly attractive potential up to a critical distance beyond which the force gets repulsive. Including multiple scattering can even lead to the appearance of a stable configuration at a large distance. Such broadband scattering forces should be observable contributions in ultra-cold atom interferometers or atomic clocks setups. They could be studied in detail in 1D~geometries with ultra-cold atoms trapped along or within an optical nanofiber. Broadband radiation force interactions might also contribute in astrophysical scenarios as illuminated cold dust clouds. 

\end{abstract}

\submitto{\NJP}
\maketitle
\section{Introduction}\label{intro}
Light scattering from point like particles is connected to momentum and energy exchange between the particles and the field. The interference of the fields scattered from different particles in an extended ensemble leads to important modifications of the total force on the ensemble~\cite{courteille2010modification,bromley2016collective} and introduces long range inter-particle light-forces~\cite{bienaime2010observation}. These forces, which originate from coherent scattering stay sizable even if the light fields are far detuned from any optical resonance~\cite{bender2010observation}. The full coupled nonlinear interaction in a cloud thus leads to a very rich and complex dynamics~\cite{douglass2012superdiffusion} including trapping, optical binding~\cite{burns1989optical,dholakia2010colloquium} and selfordering~\cite{singer2003self,tesio2012spontaneous}. Interestingly, many of the key physical effects can already be found and studied in effective 1D geometries~\cite{griesser2013light}. 
One particularly interesting example are atoms coupled to 1D~optical micro structures~\cite{zoubi2010hybrid,vetsch2010optical} as \eg an optical nanofiber, where even a single atom can significantly modify light propagation~\cite{domokos2002quantum,horak2001giant,chang2012cavity}.

In a milestone experiment, Rauschenbeutel and coworkers recently managed to trap cold atoms alongside a tapered optical nanofiber~\cite{vetsch2010optical}, a result which has been repeated and improved in several new setups~\cite{chang2012cavity,lee2013integrated,goban2012demonstration,corzo2016large,sorensen2016coherent}. As a key consequence of the strong atom-fiber coupling, optical dipole interaction and light forces between two atoms are now mediated over basically the whole fiber length. Corresponding calculations of the collective particle dynamics exhibit enhanced inter-particle forces supporting the formation of periodical self-ordered regular arrays~\cite{chang2013self,griesser2013light,holzmann2014self}, where light and motion is strongly coupled and correlated~\cite{holzmann2015collective}. 

Even for large ensembles, illumination with monochromatic laser light generates a translation invariant 1D geometry. To lowest order this implements equal strength infinite range coupling of each particle to all the others~\cite{chang2012cavity}. At sufficiently low kinetic temperature this induces crystallization of the particles and light fields with characteristic phonon excitations~\cite{chang2012cavity,griesser2013light, holzmann2015collective,ostermann2016spontaneous}. It has already been shown, that adding a second laser frequency enhances particle-particle interaction at controllable distances~\cite{Ostermann2014scattering}. Note that using circularly polarized light, asymmetric chiral scattering and interaction can be implemented in such systems as well to generate very exotic chiral spin models~\cite{pichler2015quantum}. Here for simplicity we will assume transverse linear polarization though to keep inversion symmetry.

\begin{figure}
\centering
\includegraphics[width=0.8\textwidth]{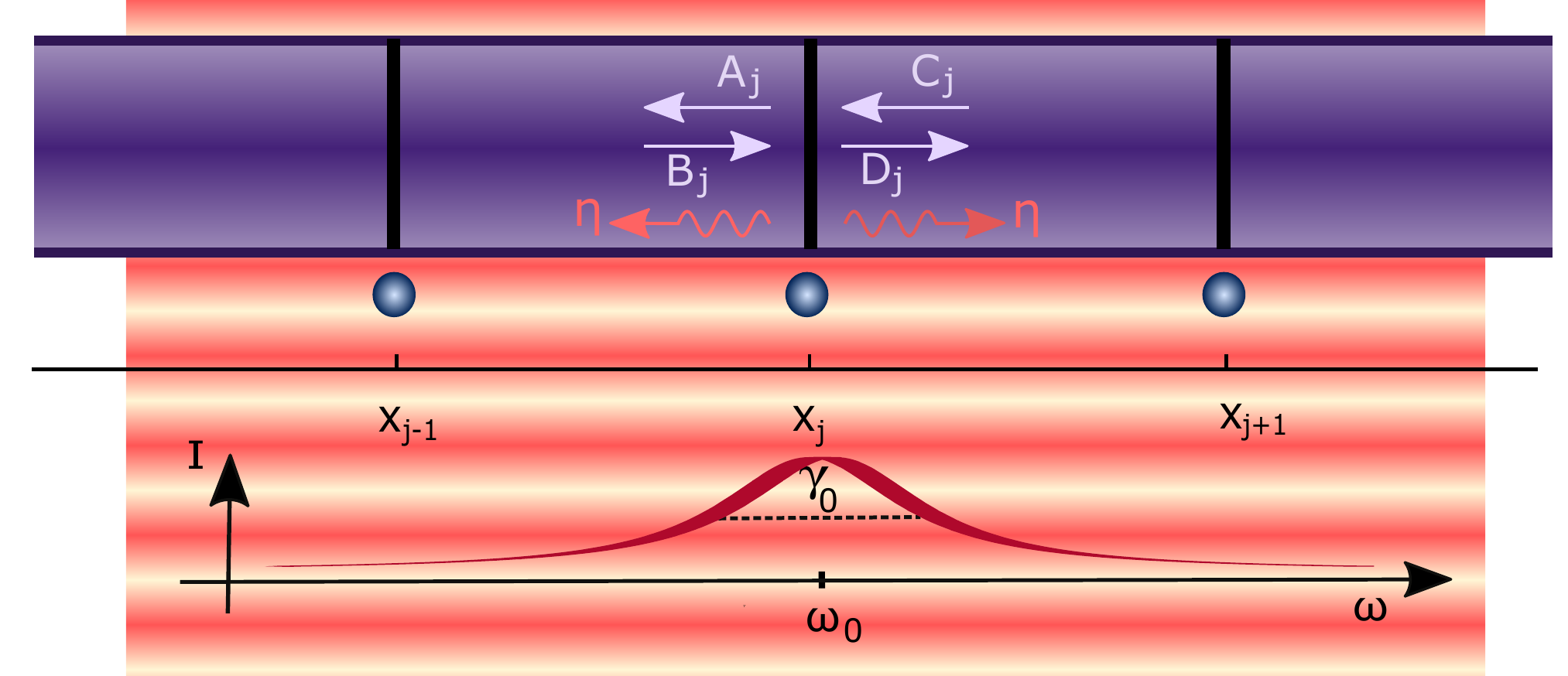}
\caption{A 1D array of point particles scattering light in and out of an optical nano-structure can be modelled as a collection of beam splitters interacting with a plane wave.}
\label{setup}
\end{figure}

In this work we generalize this coupled atom-field dynamics to incoherent light with a large bandwidth and no stable phase relation between the different frequency components. Hence the incoming radiation field is space and time translation invariant, but now possesses a finite correlation length. Surprisingly there appear substantial and partly even enhanced inter particle forces, for which we can control the effective interaction length via the spectral shape of the incoming radiation. Note that such broadband finite range forces were recently also predicted and observed in a seminal nano particle experiment with incoherent 3D illumination in a solution~\cite{brugger2015controlling}. As the corresponding full 3D model is rather complex to solve and understand, we will restrict ourselves to a quasi 1D setup as mentioned above and depicted in \fref{setup} allowing an at least partially analytic treatment. Nevertheless we expect the model to exhibit the essential underlying multiple scattering physics. Note that instead of shaping the incoming light spectrum, one can alternatively envisage to tailor the waveguide dispersion relation by nano-structuring to manipulate propagation lengths~\cite{douglas2015photon,asenjo2016atom}.  

This work is organized as follows. We first generalize the basic definitions and dynamical equations of the proven multiple scattering model for light forces in 1D systems~\cite{deutsch1995photonic, asboth2008optomechanical,sonnleitner2011optical,sonnleitner2012optomechanical} towards fields with finite bandwidth, orthogonally impinging on small particles in a 1D trap along an optical waveguide structure~\cf \fref{setup}. For two particles we analytically calculate the resulting inter-particle forces as function of bandwidth and separation. These results are then numerically extended to higher particle numbers studying the self-consistent coupled particle field dynamics for small ensembles and varying bandwidth. As a final case we will study temperature dependent forces in thermal radiation fields.   

\section{Multiple scattering approach for a point particle chain in broadband light } \label{model}
%
%
Let us assume an incoming broadband radiation field with all the wavelength components larger than the particle size and detuned from any optical resonance. The particles move in a 1D trap close enough to an optical fiber so that they can coherently scatter light into and out of the propagating fiber modes. For simplicity we assume a nanofiber, which supports only a single transverse mode propagating in each direction and a single peaked frequency distribution for the incident light field with center frequency $\omega_0$ and width $\gamma_0$. Its maximum intensity is $I_0$ so that:
\begin{equation}
I(\omega)=\frac{I_0}{\pi}\frac{\gamma_0}{\gamma_0^2+(\omega-\omega_0)^2},
\label{i0}
\end{equation}
which can be considered as a generic model of a broadband laser field or fluorescence of a large atomic ensemble. The scattered field spectrum by a single point like scatterer will than largely be proportional to the incoming distribution. For very broad light fields as generated from an LED source or ultimately using blackbody radiation with a very broad Planck spectrum, one would of course get corrections from the frequency dependent scattering amplitude $\eta$. We will only address this limit a bit at the end of this work.  

\subsection{Optical binding forces between two particles}\label{two}
To get some first insight we start with the simplest nontrivial example of two particles at a distance $d=x_2-x_1$ scattering light into the fiber. Inside the fiber the fields scattered from the two particles will interfere and a part of the field scattered by the first particle will be backscattered from the second one and vice versa. As shown in previous work, this situation can be modeled by a beam splitter approach via a $3\times 3$ coupling matrix~\cite{holzmann2014self}. Each beam splitter is parametrized by a complex polarizability $\zeta $ with dispersive part $\zeta_r$, absorptive part $\zeta_i$ and an effective scattering amplitude $\eta_0$, which for simplicity we assume to be frequency independent. In this case the multiple scattering can be summed to all orders and still allows for analytic calculations of the total self-consistent fields and forces by integrating over the whole spectrum. Note that as only the product of the spectral polarizability and the illumination spectrum appears in the calculations any frequency dependence can be absorbed into a more complex effective pump distribution.  

For two particles and small $\zeta=\zeta_r+i\zeta_i$~\cite{holzmann2014self} as well as small relative bandwidth $\gamma_0\ll\omega_0$, integration over the whole frequency range~\eref{i0} yields:
\begin{eqnarray}
\fl
\eqalign{F_1 &= \frac{I_{\eta_0}}{\pi c} \int_{-\infty}^{\infty }\frac{\gamma_0}{\gamma_0 ^2+(\omega -\omega_0)^2}\cos \left(\frac{\omega d}{c}\right) \left(1-2\zeta_i\cos\left(\frac{\omega d}{c}\right)-2\zeta_r  \sin \left(\frac{\omega d}{c}\right)\right) d\omega\\
&=\frac{I_{\eta_0}}{c}\left( e^{-\frac{\gamma_0 d}{c}}\cos\left(\frac{\omega_0 d}{c}\right)-\zeta_i\left(1+e^{-2\frac{\gamma_0 d}{c}} \cos\left(\frac{2\omega_0 d}{c}\right)\right)-\zeta_r e^{-2\frac{\gamma_0 d}{c}}\sin\left(\frac{2\omega_0 d}{c}\right) \right).}
\label{F2eq}
\end{eqnarray}

\begin{center}
\begin{figure}
\begin{minipage}{0.49\textwidth}
\includegraphics[width=\textwidth]{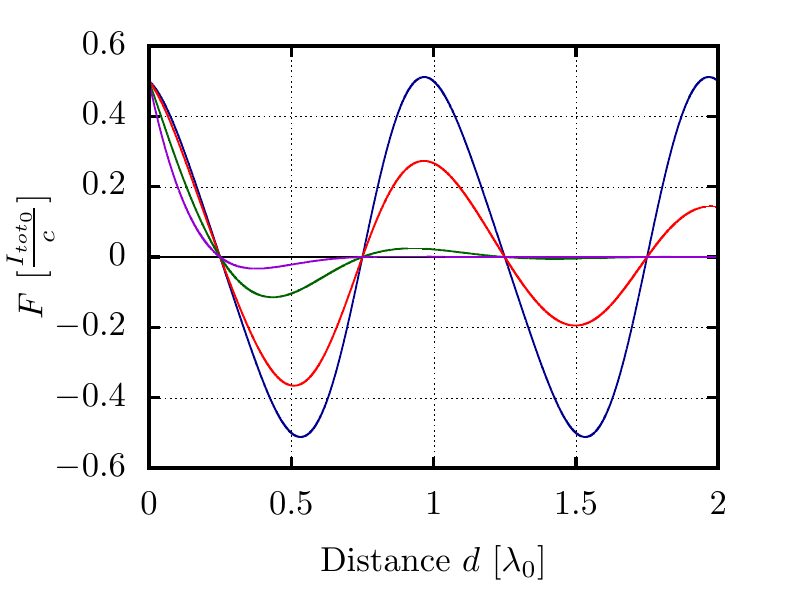}
\caption{Light force on the left of two particles as function of their distance $d$ for $\zeta=1/9$ and different bandwidths. Blue corresponds to $\gamma_0=0$, red to $\gamma_0=0.1~\omega_0$, green to $\gamma_0=0.5~\omega_0$ and violet to $\gamma_0=\omega_0 $.}
\label{F2}
\end{minipage}
\begin{minipage}{0.49\textwidth}
\vspace{-0.12\textwidth}
\includegraphics[width=\textwidth]{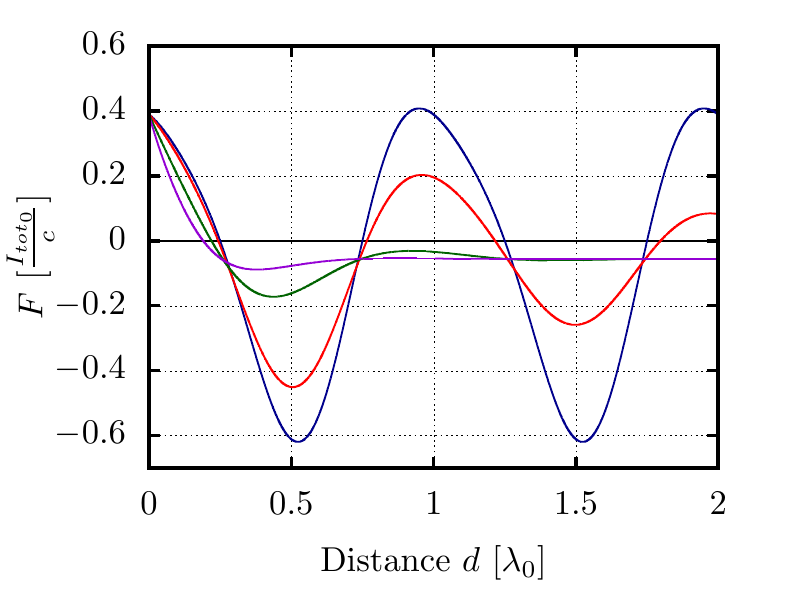}
\caption{Same as above for $\zeta=(1+i)/9$. Blue corresponds to $\gamma_0=0$, red to $\gamma_0=0.1~\omega_0$, green to $\gamma_0=0.5~\omega_0$ and violet to $\gamma_0=\omega_0 $.}
\label{F2gammazeta}
\end{minipage}
\end{figure}
\end{center}

In contrast to the infinite range for a monochromatic field~\cite{holzmann2014self} we find an additional exponential decay of the inter-particle force determined solely by the illumination bandwidth. This behavior is exhibited in figure~\ref{F2}, where the forces on the first particle for different bandwidths are compared. We still find stable configurations (optical binding) at several different distances, which can be calculated from the zero-points of the force in~\eref{F2eq}. Note that as a reference value we also have included the case of an unrealistically broad distribution of width $\gamma_0=\omega_0$ to get some qualitative insight for the infinite bandwidth limit. 

For a stable point we need zero average force and that the derivative of the force on the first beam-splitter with respect to distance is positive and negative for the second one. Surprisingly for non-absorbing particles with vanishing imaginary part (negligible absorption) of $\zeta$, $\zeta_i=0$, these stable distances $d_0$ do not depend on the bandwidth and are equal as for a monochromatic case~\cite{holzmann2014self}:

\begin{equation}
d_0=\left(\frac{3}{4}+n\right)\lambda_0,~n\in\mathbb{N}.
\end{equation}
\begin{figure}
\centering
\includegraphics[width=0.5\textwidth]{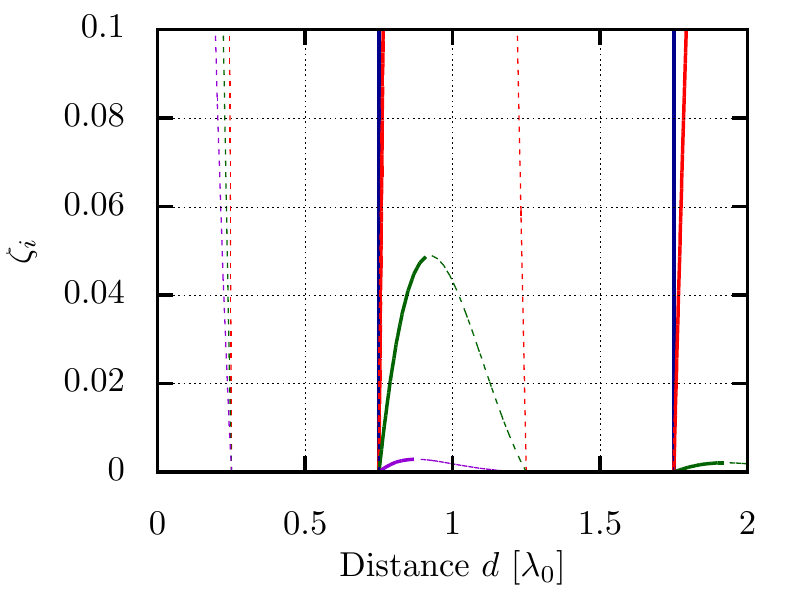}
\caption{Optical binding distances $d$ for two particles as a function of absorption coefficient $\zeta_i$ for fixed dispersive component $\zeta_r=1/9$. Blue corresponds to zero bandwidth $\gamma_0=0$, red to $\gamma_0=0.1~\omega_0$, green to $\gamma_0=0.5~\omega_0$ and violet to $\gamma_0=\omega_0 $. Solid lines show stable while dashed lines show unstable configurations.}
\label{F2gammazetai}
\end{figure}

When we also take absorption into account by introducing an imaginary part of $\zeta$, this adds extra outward radiation pressure to the force and shifts the stable distances to larger values. Eventually for too large $\zeta_i$ no stable configuration can be found as shown in figure~\ref{F2gammazeta} and~\ref{F2gammazetai}. Note that scattering of light to non propagating field modes simply appears as effective loss in the imaginary part $\zeta_i$.
\begin{center}
\begin{figure}
\begin{minipage}{0.47\textwidth}
	\centering
\includegraphics[width=\textwidth]{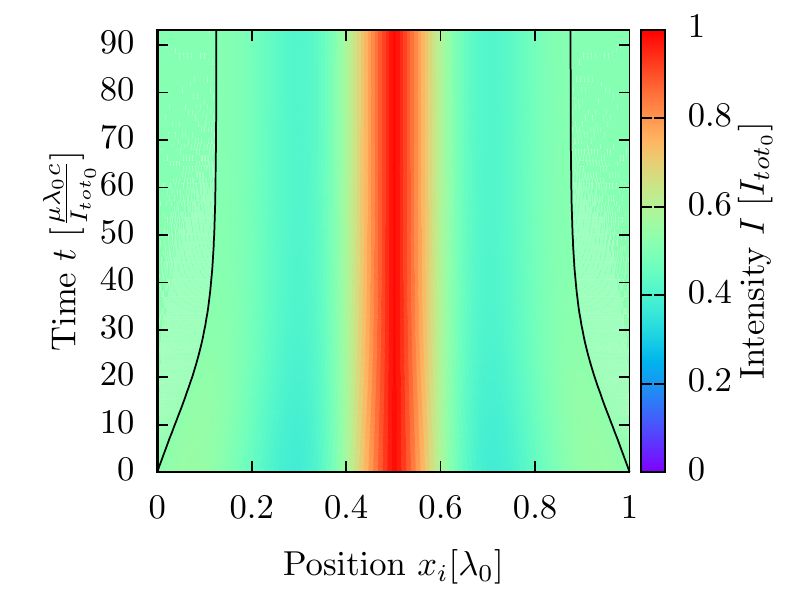}
\caption{Trajectories of two particles as a function of time for $\zeta=1/9$ and $\gamma_0=0.1~\omega_0$.}
\label{int2gammazeta1}
\end{minipage}
\hspace{0.04\textwidth}
\begin{minipage}{0.47\textwidth}
\vspace{-0.05\textwidth}	
	\centering
\includegraphics[width=\textwidth]{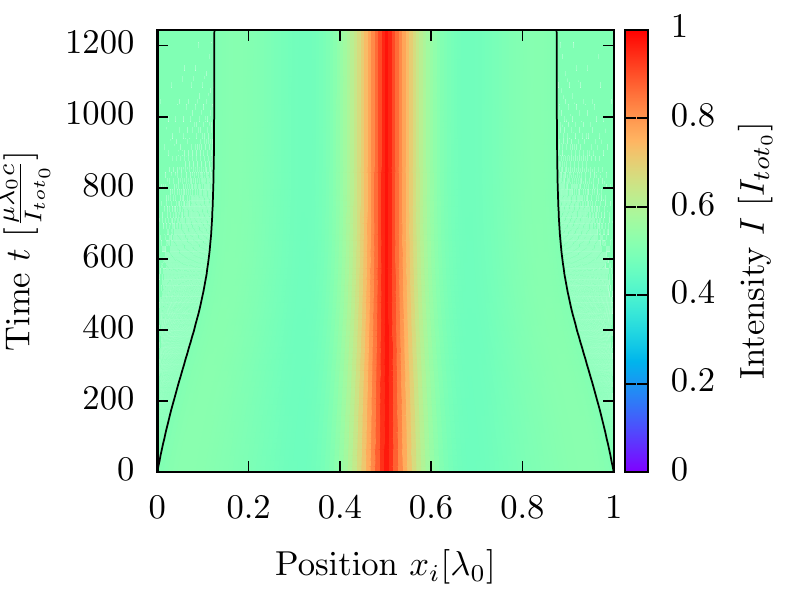}
\caption{Same as above for $\zeta=1/9$ and $\gamma_0=\omega_0$.}
\label{int2gammazeta3}
\end{minipage}
\end{figure}
\end{center}

Once knowing the forces we can also simulate the coupled particle field dynamics for non-equilibrium conditions. Adding some effective friction to particle motion we simply integrate the equations of motion and let the coupled particle field system evolve towards stable configurations. This is shown in figures~\ref{int2gammazeta1} and~\ref{int2gammazeta3}, where such trajectories for different illumination bandwidths $\gamma_0$ are compared. Indeed the simulations confirm the independence of the stable distance on the line width of the radiation. 

\begin{center}
\begin{figure}
\begin{minipage}{0.49\textwidth}
\centering
\includegraphics[width=\textwidth]{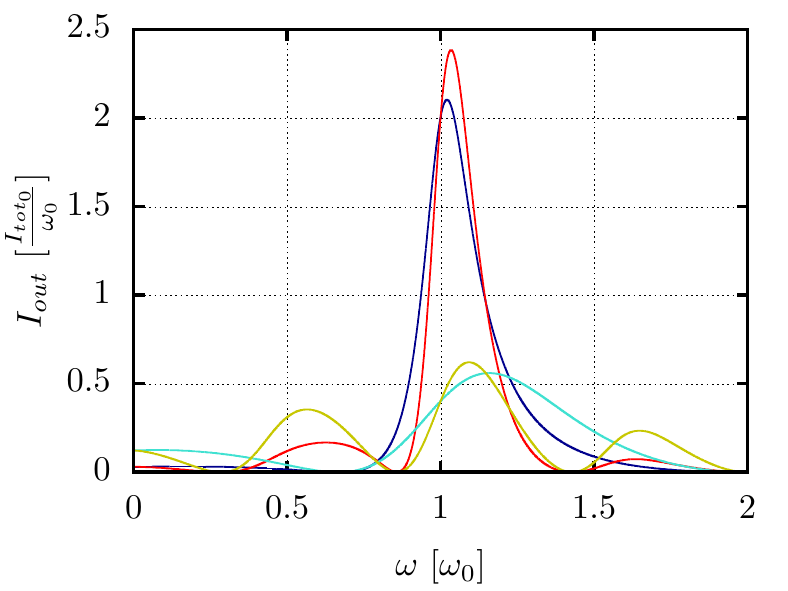}
\caption{Spectral intensity distribution of the light emitted into the fiber outside the particles for $\zeta=\frac{1}{9}$ for various pump bandwidths. Blue corresponds to $\gamma_0=0.1~\omega_0$ and $d=\frac{3}{4}~\lambda_0$, red to $\gamma_0=0.1~\omega_0$ and $d=\frac{7}{4}~\lambda_0$, cyan to $\gamma_0=0.5~\omega_0$ and $d=\frac{3}{4}~\lambda_0$, and yellow to $\gamma_0=0.5~\omega_0$ and $d=\frac{7}{4}~\lambda_0$.}
\label{Iomega}
\end{minipage}
\begin{minipage}{0.49\textwidth}
\vspace{-0.24\textwidth}
\centering
\includegraphics[width=\textwidth]{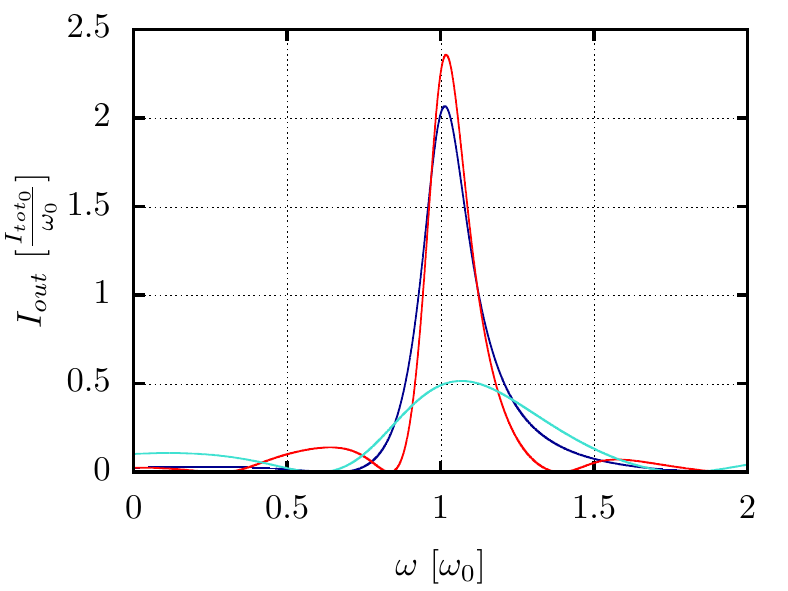}
\caption{Same as above for $\zeta=\frac{1+i}{9}$. Blue corresponds to $\gamma_0=0.1~\omega_0$ and $d=0.77~\lambda_0$, red to $\gamma_0=0.1~\omega_0$ and $d=1.8~\lambda_0$, and cyan to $\gamma_0=0.3~\omega_0$ and $d=0.83~\lambda_0$.}
\label{Iomegai}
\end{minipage}
\end{figure}
\end{center}

Note that in all cases the particles act like resonators creating a field intensity maximum in between them, while they themselves do not sit at either a field maximum or minimum as one might expect from simple light shift considerations. As a clear signature of multiple scattering we also get a strong spectral filtering response reminiscent of a Fabry Perot cavity from our particle pair. While some frequencies are strongly confined, others are dominantly transmitted depending on the particle distance. As a consequence the spectral distribution of the intensity emitted into the fiber outside the particle pair is strongly modified from the input. It also substantially changes with varying polarizability $\zeta$ and the line width $\gamma_0$.

This is depicted in some typical examples in figure~\ref{Iomega} and \ref{Iomegai}. Unexpectedly the intensities are not spectrally symmetric, which is a consequence of the strong frequency-dependent scattering of the beam-splitters. In this case some frequencies are filtered by particles, while others can pass them. Which frequencies can pass does not only depend on $\zeta$ and $\gamma_0$ but also at which of the stable distances the particles order. Thus, one can directly get information on the distance of the particles by measuring the frequency distribution of the outgoing intensity.

\subsection{Forces and selfordering for several beam splitters}

In principle determining the field distributions, the forces and stationary states for a larger number of beam splitters is straightforward by multiplying the corresponding scattering and propagation matrices. In practice, useful analytic results can only be obtained by ignoring backscattering and setting $\zeta=0$ as multiple scattering leads to very complex expressions. In contrast to the monochromatic case the particles in steady state here are not equidistantly distributed, which leads to a much more complex transfer matrix. Nevertheless, for $\zeta=0$ the transfer matrix still can be calculated analytically for $N$ particles and integrated over the frequency distribution to yield the following expression for the total force:
\begin{eqnarray}
\fl
\eqalign{F_{j,N}&=\frac{I_{\eta_0}}{c\pi}\int_{-\infty}^{\infty} d\omega \frac{\gamma_0}{\gamma_0^2+(\omega-\omega_0)^2}\left(\sum _{l=j}^{N-1} \cos \left(\frac{\omega\sum _{i=j}^l d_i}{c}\right)-\sum _{l=1}^{j-1} \cos \left(\frac{\omega\sum _{i=l}^{j-1} d_i}{c}\right)\right)\\
&=\frac{I_{\eta_0}}{c}\left(\sum _{l=j}^{N-1}e^{-\sum _{i=j}^l\frac{\gamma_0d_i}{c}} \cos \left(\frac{\omega\sum _{i=j}^l d_i}{c}\right)-\sum _{l=1}^{j-1} e^{-\sum _{i=l}^{j-1}\frac{\gamma_0d_i}{c}}\cos \left(\frac{\omega\sum _{i=l}^{j-1} d_i}{c}\right)\right).}
\label{FomegaN}
\end{eqnarray}
\begin{center}
\begin{figure}
\begin{minipage}{0.49\textwidth}
	\centering
\includegraphics[width=\textwidth]{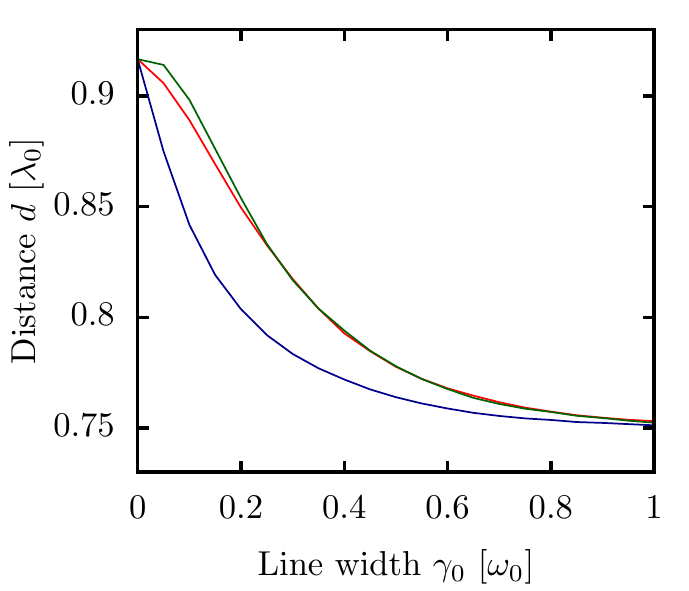}
\caption{Shortest stable distances for a symmetric configuration of six particles as a function of bandwidth $\gamma_0$. Blue corresponds to the distance between the first two particles, red to the distance between the second and the third and green to the distance between the third and the fourth particle.}
\label{gammad}
\end{minipage}
\begin{minipage}{0.49\textwidth}
	\centering
\includegraphics[width=\textwidth]{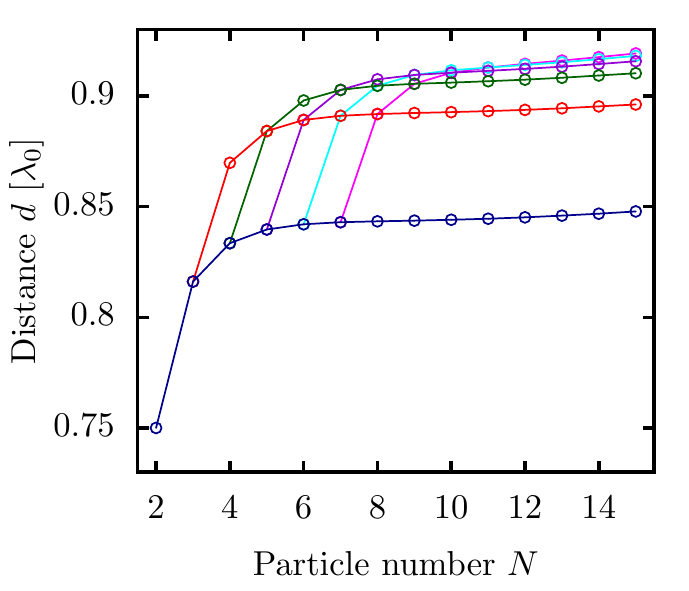}
\caption{Shortest stable distances as a function of the number of particles $N$ for $\gamma=0.1~\omega_0$. The blue line corresponds to the distance between the first two particles $d_1$, red to $d_2$, green to $d_3$, violet to $d_4$, light blue to $d_5$ and magenta to $d_6$.}
\label{FNstable}
\end{minipage}
\label{symtraj}
\end{figure}
\end{center}

As the zeros of this function yielding the stable configurations are determined by a transcendent equation they can not be calculated analytically. Nevertheless a numerical solution requires very little effort. So in figure~\ref{gammad} we plot the closed configuration of stationary distances for six particles as a function of the bandwidth $\gamma_0$. Interestingly the particles are only equidistantly ordered in the two extreme limits of zero bandwidth $\gamma_0=0$ and a very large bandwidth, where interactions are short distance. For finite $\gamma_0$ the distances between the inner particles are larger than the distances between the outer ones.

For large bandwidth $\gamma_0$ the particles only significantly interact with their direct neighbors yielding equidistant order at a distance $d\approx\frac{3}{4}~\lambda_0$. This is the same distance as for two particles as a consequence of the short range of the force. Note that by controlling the input bandwidth we thus can switch from infinite range to nearest neighbor interaction. \\
This behavior is also shown in figure~\ref{FNstable}, where the dependence of the stable distance as function of the number of particles is compared for different bandwidth $\gamma_0$. The distance between the particles for larger $\gamma_0$ is closer to the two-particle-distance $\frac{3}{4}~\lambda_0$. Additionally they again confirm that the outer particles are closer than the inner ones and that the distances for the outer ones become more and more equal for large particle numbers.
\begin{figure}
\begin{minipage}{0.5\textwidth}
\centering
\includegraphics[width=\textwidth]{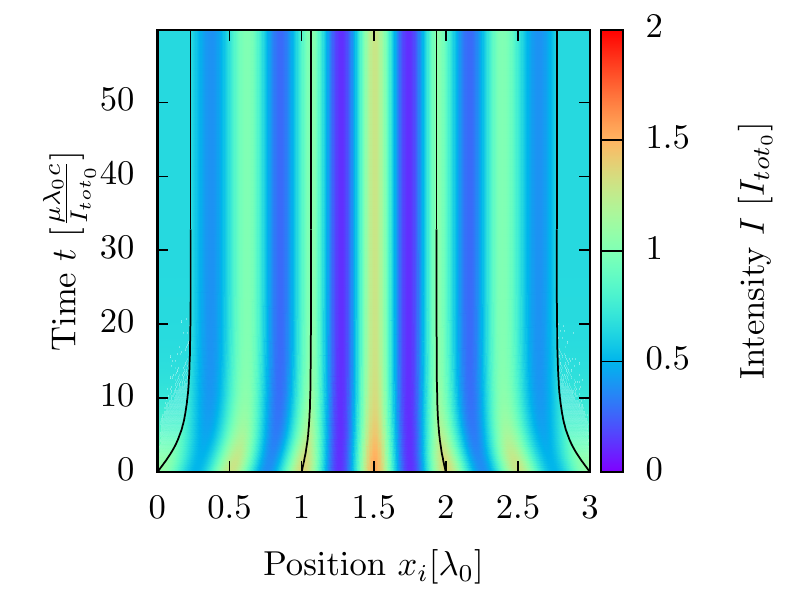}
\caption{Time evolution of atom-field distribution for four particles for $\zeta=0$ and initial distance $\gamma_0=0.1~\omega_0$.}
\label{timeevolution4a}
\end{minipage}
\begin{minipage}{0.5\textwidth}
\vspace{-0.05\textwidth}
\centering
\includegraphics[width=\textwidth]{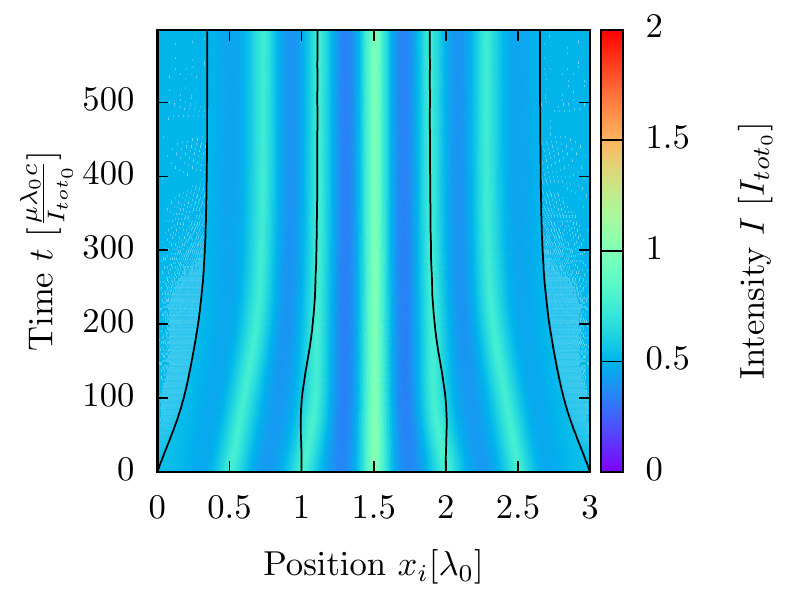}
\caption{Same as above for $\zeta=0$ and initial distance $\gamma_0=0.5~\omega_0$.}
\label{timeevolution4b}
\end{minipage}
\end{figure}

Again the couple atom-field dynamics shows intriguing physics. Figure~\ref{timeevolution4a} and \ref{timeevolution4b} show two examples for the time evolution of four particles in a broadband field. Again the distance between the outer particles is larger than between the inner particles. Interestingly the particles tend to create high field intensities between them, with the outer two providing trap sites for the inner pair and a total field maximum generated at the center. This effect, more prominent for smaller bandwidth, reminds of a self assembled cavity QED system. 

\section{Interparticle forces induced by blackbody radiation}
\label{blackbody}

As a final and extreme limit of broadband radiation let us consider the particles to be illuminated by a thermal radiation field (BBR) with a Planck spectrum. At temperature $T$ with Boltzmann constant $k_B$, Planck constant $\hbar$ and light velocity $c$ the spectral distribution reads:
\begin{equation}
I_0(\omega)=\frac{\hbar}{2\pi^2 c^2}\frac{\omega^3}{e^{\frac{\hbar\omega}{k_B T}}-1}.
\label{bbint}
\end{equation}
Of course the assumption of constant linear polarizability is only approximately valid throughout the whole Planck spectrum. Nevertheless it has been shown recently, that for atoms with a first excited state in the UV region as e.g. atomic hydrogen, this approximation is surprisingly good up to temperatures of $T\approx 6000~K$~\cite{sonnleitner2013attractive, brugger2015controlling}. Introducing a more realistic frequency dependent polarizability and resonances will introduce some quantitative changes, but one can expect qualitative agreement as long as the majority of the incident power is sufficiently detuned from resonances. This is generally the case for {\it low temperature} radiation with a peak below the visible. 

It has been shown before that the emitted BBR field from a hot source itself induces a surprising attractive force pulling atoms towards the hot object~\cite{sonnleitner2013attractive}. Here we start with two temperature less particles illuminated from the side by a BBR field. Again we integrate the fiber mediated scattering force between them over the BBR frequency range in~\eref{bbint}, which still can be done analytically to obtain 
\begin{equation}
\label{bbr2}
F_1=-F_2=\frac{\hbar\omega_T^4}{2\pi^2 c^3}\left(\frac{\cosh\left(\frac{2d}{r_T}\right)+2}{\sinh^4\left(\frac{d}{r_T}\right)}-3\left(\frac{r_T}{d}\right)^4\right),
\end{equation}

where we defined a thermal length $r_T=\frac{\hbar c}{\pi k_B T}$ and a thermal frequency $\omega_T=\frac{\pi k_B T}{\hbar}$. For particles in an environment with room temperature the thermal length gives $r_{300K}\approx \frac{7.3\cdot 10^{-4}}{T}~m\cdot K\approx 2.4\cdot 10^{-6}~m$, which is much larger then a typical optical wavelength. Of course the long wavelength components will not be significantly guided by the fiber, but due to their low energy and momentum content, these parts are not relevant for the resulting force anyway.

\begin{figure}
\centering
\includegraphics[width=0.5\textwidth]{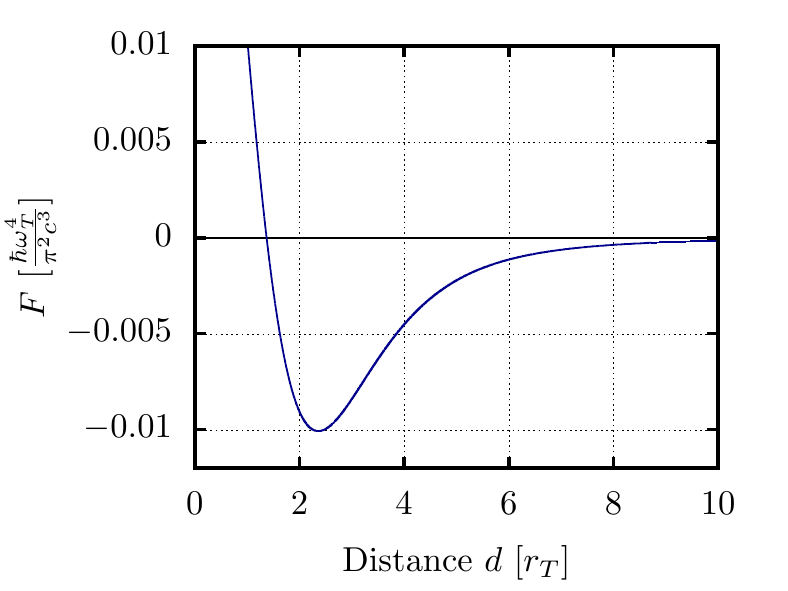}
\caption{Force on the first of two particles in a blackbody radiation field as a function of the distance for $\zeta=0$.}
\label{F2bb}
\end{figure}

\begin{figure}
\begin{minipage}{0.49\textwidth}
\centering
\includegraphics[width=\textwidth]{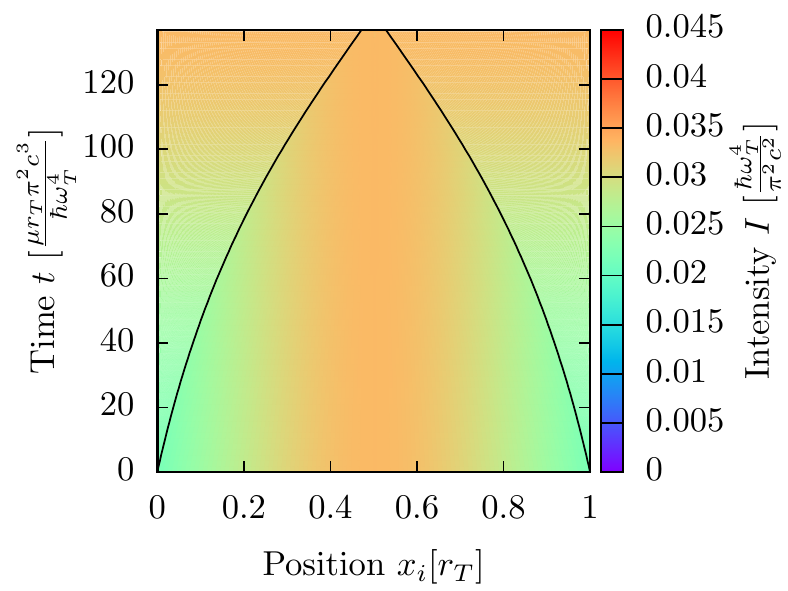}
\caption{Trajectories of two particles at $\zeta=0$ with initial distance $d=\lambda_0$.}
\label{intposTpdf}
\end{minipage}
\begin{minipage}{0.49\textwidth}
\centering
\includegraphics[width=\textwidth]{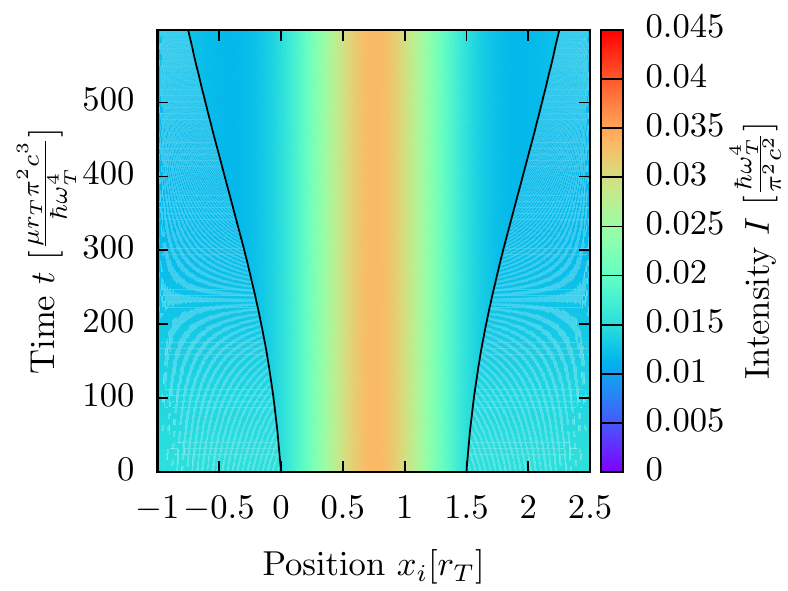}
\caption{Same as above for $\zeta=0$ with initial distance $d=1.5~\lambda_0$.}
\label{intposT2pdf}
\end{minipage}
\end{figure}

From~\eref{bbr2} we can expect a distance and temperature dependent sign change of this force including a zero force distance. In fact as depicted in figure~\ref{F2bb}, the particles will attract each other if they are close but very surprisingly this changes to repulsion at distances larger than the thermal radius $r_T$. We thus find no stable equilibrium distance. The strongest force appears at short distances and will induce a capture range below which two particles will collapse together. This will large determine the effective scattering size of two particles. As said, the zero force equilibrium point ($d_0=1.37~r_T$) is not stable and a small perturbation causes collapse or separation. This behavior is plotted in figure~\ref{intposTpdf} and \ref{intposT2pdf}, where in the first case the particles are attracted by the high intensity peak forming between them, while they are repelled in the second case.

While the quantitative accuracy of this prediction can be doubted such a qualitative behavior should be generally true, as 3D calculations yield similar attractive behavior~\cite{brugger2015controlling}. This could imply important changes in the dynamics of ultra cold particle ensembles in thermal radiation fields with even astrophysical consequences.

\begin{figure}
\begin{minipage}{0.49\textwidth}
\centering
\includegraphics[width=\textwidth]{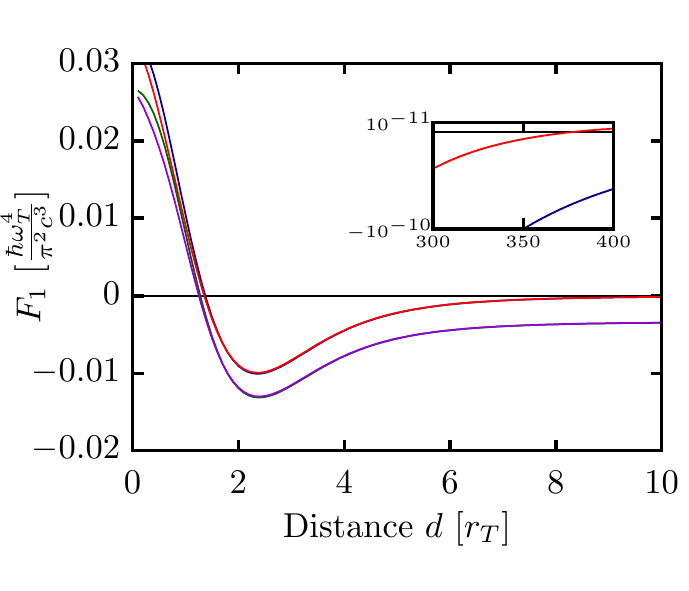}
\vspace{-0.85cm}
\caption{Force on the first of two particles as a function of the distance. Blue corresponds to $\zeta=0$, red to $\zeta=0.1$, green to $\zeta=0.1i$ and violet to $\zeta=0.1(1+i)$.}
\label{F2zeta}
\end{minipage}
\begin{minipage}{0.49\textwidth}
\vspace{-0.55cm}
\centering
\includegraphics[width=\textwidth]{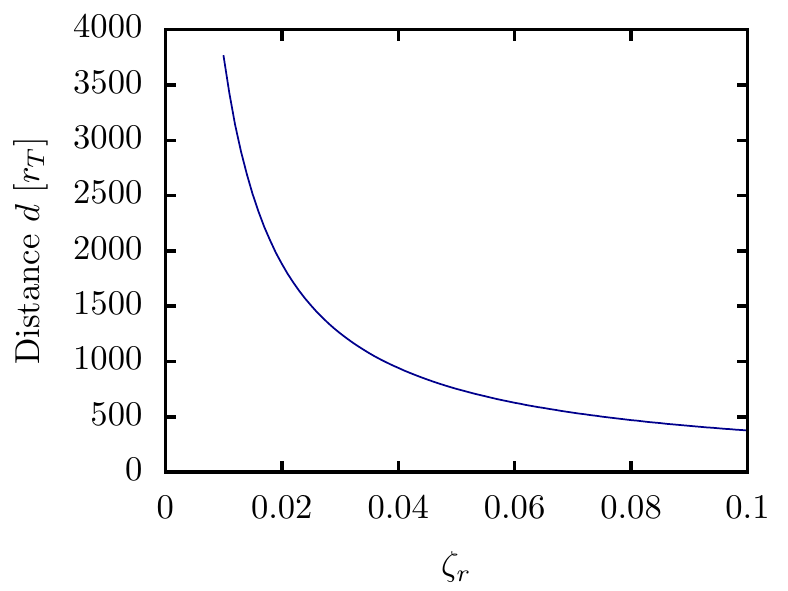}
\caption{Stable zero-points of the force on two particles as a function of the distance $d$ and $\zeta_r$.}
\label{F2zeta0}
\end{minipage}
\end{figure}

Interestingly, the physics changes when we include backscattering and absorption of the radiation by the particles via introducing a finite $\zeta$-parameter. Figure~\ref{F2zeta} and~\ref{F2zeta0} show that for real $\zeta>0$ we get long range attraction and can even find a second zero point of the force at a large distance. As this effect is very tiny it might be more of academic than practical interest, but this separation is even stable against perturbations. Including absorption via imaginary parts of $\zeta$ has the opposite effect on the particles as expected. The force is modified in a way that in the limit $\zeta_r\ll 1$ no stable configurations can be found due to long range radiation pressure contributions.
\begin{figure}
\centering
\includegraphics[width=0.5\textwidth]{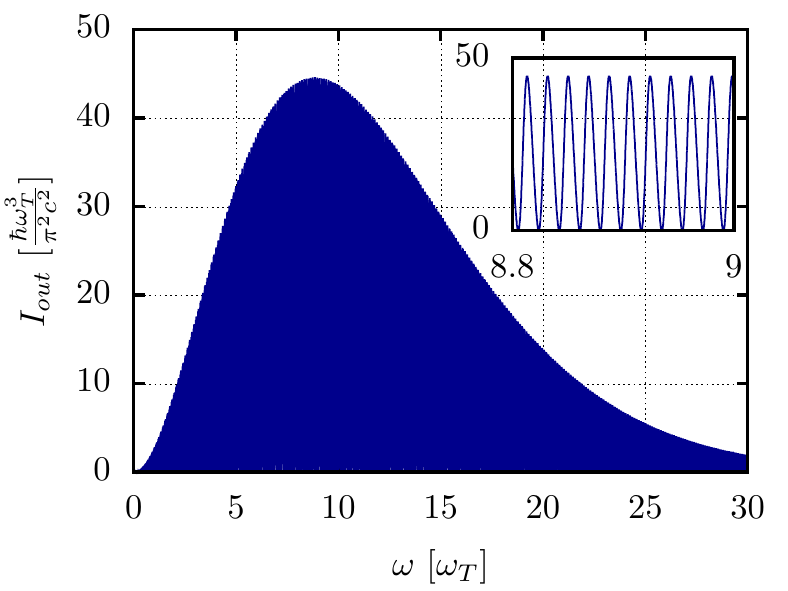}
\caption{Outgoing intensity as a function of $\omega$ for $\zeta=\frac{1}{9}$ at the stable position $d_0=377~r_T$.}
\label{Iomegabb}
\end{figure}

Similar to the previous case the particle also filter the light and sculpt an oscillating transmission pattern into the outgoing light. Figure~\ref{Iomegabb} shows the outgoing intensities for a blackbody radiation field, which keep the gross shape of BBR but again the particles tend to filter some frequencies.

\subsection{Larger ensembles in BBR fields}

\begin{figure}
\begin{minipage}{0.49\textwidth}
\centering
\includegraphics[width=\textwidth]{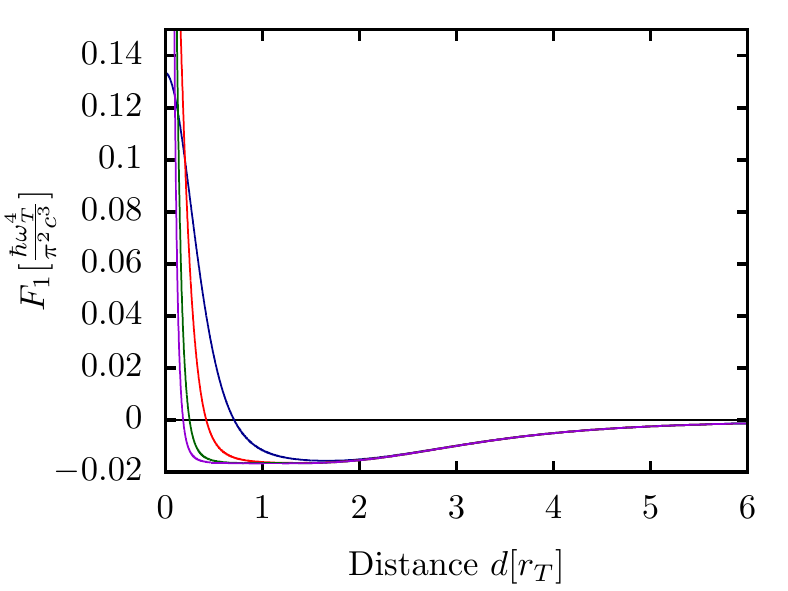}
\caption{Force on the first of 5 (blue), 10 (red), 20 (green), 30 (violet) particles as a function of the distance for $\zeta=0$.}
\label{Fnbb}
\end{minipage}
\begin{minipage}{0.49\textwidth}
\vspace{0.05\textwidth}
\centering
\includegraphics[width=\textwidth]{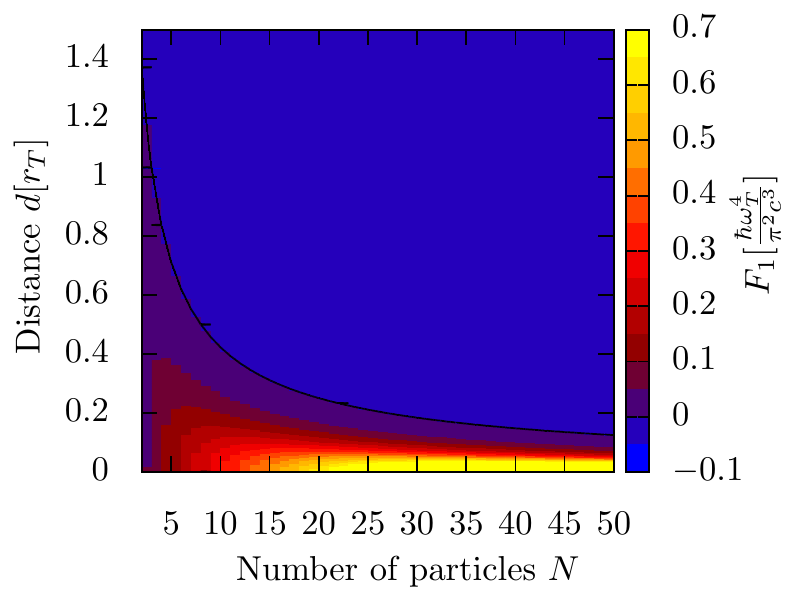}
\caption{Contour-plot of the force as a function of the particle number $N$ and the distance $d$ for $\zeta=0$. The black line corresponds to the zero-force line.}
\label{NDbb}
\end{minipage}
\end{figure}

\begin{figure}
\centering
\includegraphics[width=0.5\textwidth]{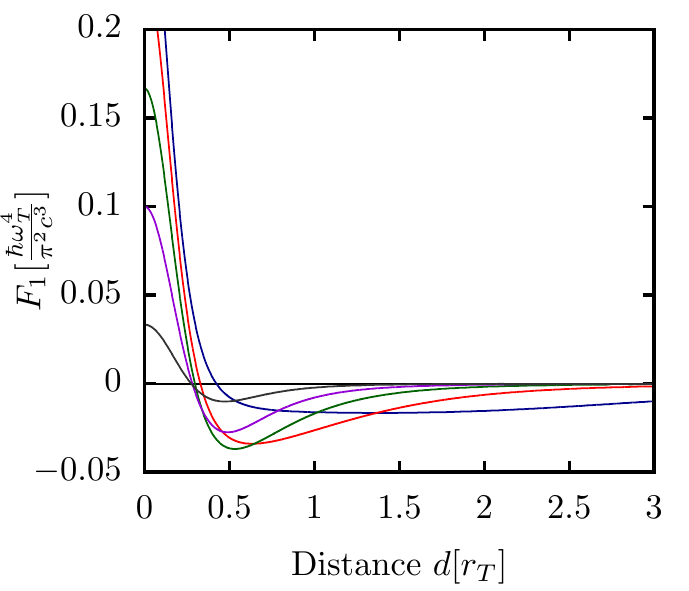}
\caption{Force on the first (blue), second (red), third (green), fourth (violet), fifth (grey) of ten particles as a function of the distance for $\zeta=0$.}
\label{F10bb}
\end{figure}

\begin{figure}
\begin{minipage}{0.49\textwidth}
\centering
\includegraphics[width=\textwidth]{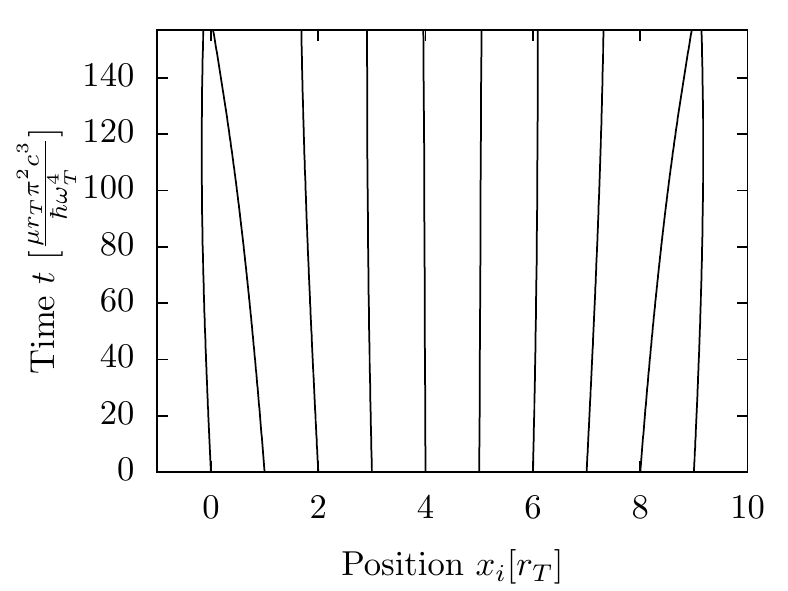}
\caption{Time evolution for ten particles for $\zeta=0$ and initial distance $d=\lambda_0$.}
\label{pos10bb}
\end{minipage}
\vspace{0.1\textwidth}
\begin{minipage}{0.49\textwidth}
\centering
\includegraphics[width=\textwidth]{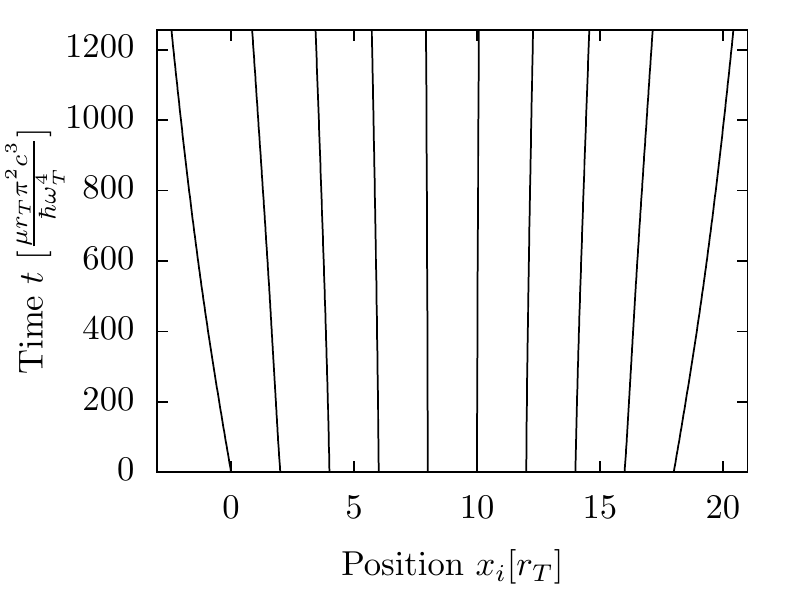}
\caption{Same as above for $\zeta=0$ and initial distance $d=2~\lambda_0$.}
\label{pos10bb2}
\end{minipage}
\end{figure}

\begin{figure}
\begin{minipage}{0.49\textwidth}
\centering
\includegraphics[width=\textwidth]{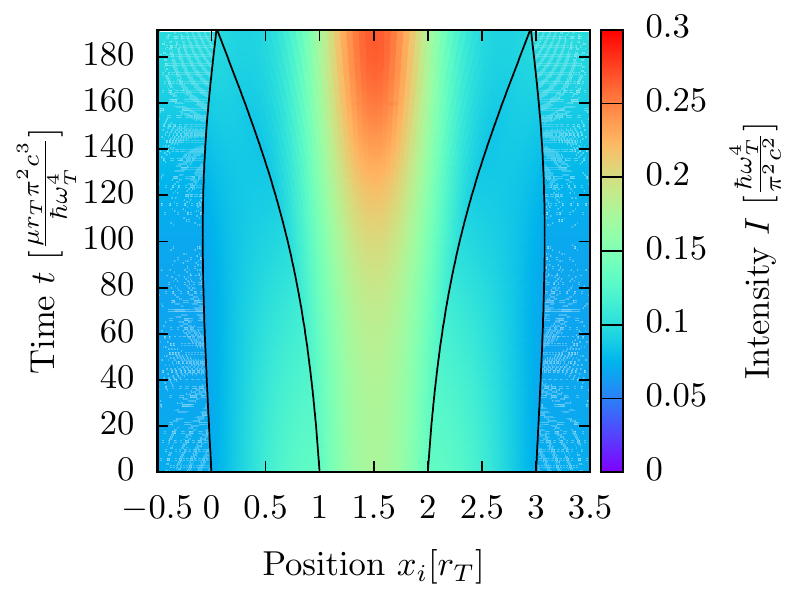}
\caption{Time evolution for four particles for $\zeta=0$ and initial distance $d=\lambda_0$.}
\label{pos4bba}
\end{minipage}
\begin{minipage}{0.49\textwidth}
\centering
\includegraphics[width=\textwidth]{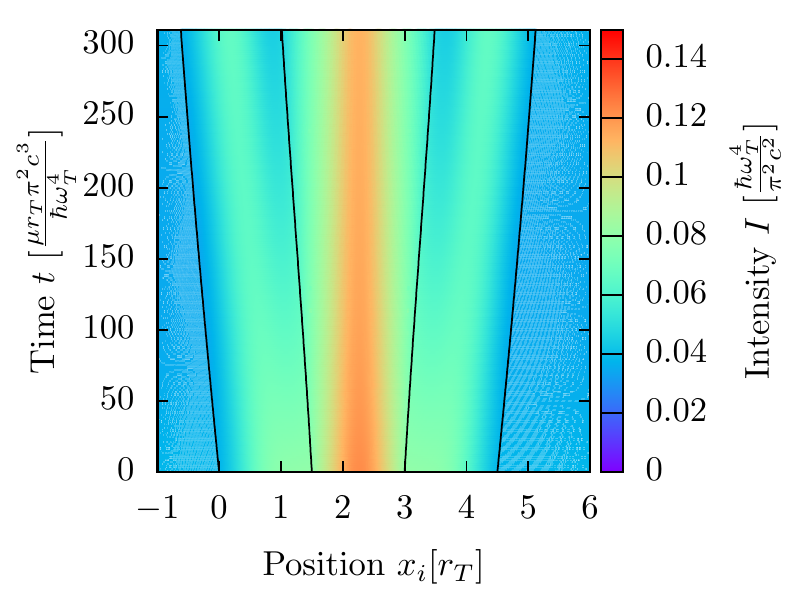}
\caption{Same as above for $\zeta=0$ and initial distance $d=1.5~\lambda_0$.}
\label{pos4bbb}
\end{minipage}
\end{figure}

BBR induced forces on larger ensembles can be calculated in the same way as before. As expected this leads to the rather complex analytic expression in terms of simple sums for the forces that govern the dynamics of the cloud, but allows simple numerical evaluation:
\begin{eqnarray}
\fl
\eqalign{F_\Omega&=\frac{\hbar}{2\pi^2c^3}\int_0^\infty d\omega \frac{\omega^3}{e^{\frac{\hbar\omega}{k_B T}}-1}\left(\sum _{l=j}^{N-1} \cos \left(k \sum _{i=j}^l d_i\right)-\sum _{l=1}^{j-1} \cos \left(k\sum _{i=l}^{j-1} d_i\right)\right)\\
&=\frac{\hbar\omega_T^4}{2\pi^2c^3}\left(\sum_{l=j}^{N-1}\left(\frac{\cosh \left(\frac{2\sum_{i=j}^l d_i}{r_T}\right)+2}{\sinh^4\left(\frac{\sum_{i=j}^l d_i}{r_T}\right)}-\frac{3 r_T^4}{\left(\sum_{i=j}^l d_i\right)^4}\right)\right.\\
&\left.-\sum_{l=1}^{j-1}\left(\frac{\cosh \left(\frac{2 \sum_{i=l}^{j-1}d_i}{r_T}\right)+2}{\sinh^4\left(\frac{\sum_{i=l}^{j-1}d_i}{r_T}\right)}-\frac{3r_T^4}{\left(\sum_{i=l}^{j-1}d_i\right)^4}\right)\right).}
\label{Fnbbeq}
\end{eqnarray}
As it is rather hopeless to analytically find the zeros of this function, in order to get some first insight in figure~\ref{Fnbb} we simply plot the force on the outermost particle as function of the distance for an equidistant particle array and weak backscattering amplitude. Note that the magnitude of the near field attraction strongly increases with the particle number, while its range decreases at the same time. Hence this would trigger a fast collapse of a particle cloud once a certain small distance (high density) has been reached. At larger distances the force on the first particle does not depend on the total particle number and is repulsive. This effect is also visible in figure~\ref{NDbb}.
Figure~\ref{F10bb}, which shows the force on the first five of ten particles clearly shows that the force on the inner particles is smaller than on the outer ones. Only for some smaller distances this is not true, which is also shown in figure~\ref{pos10bb}, while we choose a larger distance in figure~\ref{pos10bb2}. The simulations confirm that the particles do not form stable configurations but tend either to collapse together or repulse each other towards infinity. This effect can also be observed for four particles in figure~\ref{pos4bba} and~\ref{pos4bbb}. For this case large intensities between the particles push them apart.
Here definitely more extensive and detailed simulations would be needed to fully understand the dynamics, which, however, is beyond the scope of this work. 

%
\section{Conclusions}
We have shown that fiber mediated broadband light scattering from linear arrays of polarizable particles induces a much more rich and complex inter particle forces than simple repulsive radiation pressure. In fact one can tailor the interaction from infinite range to nearest neighbor coupling by a suitable choice of the bandwidth. This originates from the frequency and distance dependent interference of the various frequency components in the field, which appears for dispersive as well as for absorptive scattering. For a wide range of bandwidths, one finds multi-stable regular ordering of the particles and long range forces as for optical binding in monochromatic fields.
The range of inter particle interactions can be well tuned via the bandwidth of the incoming field. While we have only studied single peaked frequency distributions to limit interaction ranges, multiple peaks can be expected to allow for a much more general design of controlling interaction strength with distance. The physical effect of these generally quite weak interactions can be significantly enhanced and studied in a controlled way by coupling via optical nanostructures as nanofibers. Some preliminary calculations, nevertheless, show that qualitatively analogous behavior survives in 2D and 3D situations.

Surprisingly even for very broadband BBR fields one finds multi particle collective effects and a complex interplay of attraction and repulsion at distances around the thermal radius, which will lead to a complex nonlinear dynamics and response of larger ensembles. While this certainly is hard to observe in standard setups~\cite{brugger2015controlling} it still could have significant influence on longer time scales. Alternatively it could be measurable in very force sensitive setups such as atom interferometers where it should induce extra density dependent shifts and dephasing. 

\ack
We thank Juan Jos{\'e} S{\'a}enz, Holger Muller and  Tobias Griesser for stimulating discussions. We  acknowledge support via the Austrian Science Fund grant SFB F4013 . 
%

%


\section*{References}
\begin{thebibliography}{10}
\expandafter\ifx\csname url\endcsname\relax
  \def\url#1{{\tt #1}}\fi
\expandafter\ifx\csname urlprefix\endcsname\relax\def\urlprefix{URL }\fi
\providecommand{\eprint}[2][]{\url{#2}}

\bibitem{courteille2010modification}
Courteille P~W, Bux S, Lucioni E, Lauber K, Bienaime T, Kaiser R and Piovella N
  2010 {\em The European Physical Journal D\/} {\bf 58} 69--73

\bibitem{bromley2016collective}
Bromley S~L, Zhu B, Bishof M, Zhang X, Bothwell T, Schachenmayer J, Nicholson
  T~L, Kaiser R, Yelin S~F, Lukin M~D {\em et~al.\/} 2016 {\em Nature
  communications\/} {\bf 7}

\bibitem{bienaime2010observation}
Bienaime T, Bux S, Lucioni E, Courteille P~W, Piovella N and Kaiser R 2010 {\em
  Physical review letters\/} {\bf 104} 183602

\bibitem{bender2010observation}
Bender H, Stehle C, Slama S, Kaiser R, Piovella N, Zimmermann C and Courteille
  P~W 2010 {\em Physical Review A\/} {\bf 82} 011404

\bibitem{douglass2012superdiffusion}
{Douglass} K~M, {Sukhov} S and {Dogariu} A 2012 {\em Nature Photonics\/} {\bf
  6} 834--837

\bibitem{burns1989optical}
Burns M~M, Fournier J~M and Golovchenko J~A 1989 {\em Physical Review
  Letters\/} {\bf 63} 1233

\bibitem{dholakia2010colloquium}
Dholakia K and Zem{\'a}nek P 2010 {\em Reviews of modern physics\/} {\bf 82}
  1767

\bibitem{singer2003self}
Singer W, Frick M, Bernet S and Ritsch-Marte M 2003 {\em J. Opt. Soc. Am. B\/}
  {\bf 20} 1568--1574

\bibitem{tesio2012spontaneous}
Tesio E, Robb G, Ackemann T, Firth W and Oppo G~L 2012 {\em Physical Review
  A\/} {\bf 86} 031801

\bibitem{griesser2013light}
Grie{\ss}er T and Ritsch H 2013 {\em arXiv preprint arXiv:1303.7359\/}

\bibitem{zoubi2010hybrid}
Zoubi H and Ritsch H 2010 {\em New Journal of Physics\/} {\bf 12} 103014

\bibitem{vetsch2010optical}
Vetsch E, Reitz D, Sagu{\'e} G, Schmidt R, Dawkins S and Rauschenbeutel A 2010
  {\em Physical review letters\/} {\bf 104} 203603

\bibitem{domokos2002quantum}
Domokos P, Horak P and Ritsch H 2002 {\em Physical Review A\/} {\bf 65} 033832

\bibitem{horak2001giant}
Horak P, Domokos P and Ritsch H 2001 {\em arXiv preprint quant-ph/0108006\/}

\bibitem{chang2012cavity}
Chang D, Jiang L, Gorshkov A and Kimble H 2012 {\em New Journal of Physics\/}
  {\bf 14} 063003

\bibitem{lee2013integrated}
Lee J, Park D, Mittal S, Dagenais M and Rolston S 2013 {\em arXiv preprint
  arXiv:1303.2922\/}

\bibitem{goban2012demonstration}
Goban A, Choi K, Alton D, Ding D, Lacro{\^u}te C, Pototschnig M, Thiele T,
  Stern N and Kimble H 2012 {\em Physical Review Letters\/} {\bf 109} 33603

\bibitem{corzo2016large}
{Corzo} N~V, {Gouraud} B, {Chandra} A, {Goban} A, {Sheremet} A~S, {Kupriyanov}
  D~V and {Laurat} J 2016 {\em ArXiv e-prints\/} (\textit{Preprint}
  \eprint{1604.03129})

\bibitem{sorensen2016coherent}
S{\o}rensen H, B{\'e}guin J~B, Kluge K, Iakoupov I, S{\o}rensen A, M{\"u}ller
  J, Polzik E and Appel J 2016 {\em arXiv preprint arXiv:1601.04869\/}

\bibitem{chang2013self}
Chang D~E, Cirac J~I and Kimble H~J 2013 {\em Phys. Rev. Lett.\/} {\bf 110}
  113606

\bibitem{holzmann2014self}
{Holzmann} D, {Sonnleitner} M and {Ritsch} H 2014 {\em Eur. Phys. J. D\/} {\bf
  68} 352

\bibitem{holzmann2015collective}
Holzmann D and Ritsch H 2015 {\em Opt. Express\/} {\bf 23} 31793--31806

\bibitem{ostermann2016spontaneous}
Ostermann S, Piazza F and Ritsch H 2016 {\em Phys. Rev. X\/} {\bf 6}(2) 021026

\bibitem{Ostermann2014scattering}
Ostermann S, Sonnleitner M and Ritsch H 2014 {\em New Journal of Physics\/}
  {\bf 16} 043017

\bibitem{pichler2015quantum}
Pichler H, Ramos T, Daley A~J and Zoller P 2015 {\em Physical Review A\/} {\bf
  91} 042116

\bibitem{brugger2015controlling}
Br{\"u}gger G, Froufe-P{\'e}rez L~S, Scheffold F and S{\'a}enz J~J 2015 {\em
  Nature communications\/} {\bf 6}

\bibitem{douglas2015photon}
Douglas J~S, Caneva T and Chang D~E 2015 {\em arXiv preprint
  arXiv:1511.00816\/}

\bibitem{asenjo2016atom}
Asenjo-Garcia A, Hood J, Chang D and Kimble H 2016 {\em arXiv preprint
  arXiv:1606.04977\/}

\bibitem{deutsch1995photonic}
Deutsch I, Spreeuw R, Rolston S and Phillips W 1995 {\em Physical Review A\/}
  {\bf 52} 1394

\bibitem{asboth2008optomechanical}
Asb{\'o}th J, Ritsch H and Domokos P 2008 {\em Physical Review A\/} {\bf 77}
  063424

\bibitem{sonnleitner2011optical}
Sonnleitner M, Ritsch-Marte M and Ritsch H 2011 {\em EPL (Europhysics
  Letters)\/} {\bf 94} 34005

\bibitem{sonnleitner2012optomechanical}
Sonnleitner M, Ritsch-Marte M and Ritsch H 2012 {\em New Journal of Physics\/}
  {\bf 14} 103011

\bibitem{sonnleitner2013attractive}
Sonnleitner M, Ritsch-Marte M and Ritsch H 2013 {\em Physical review letters\/}
  {\bf 111} 023601

\end{thebibliography}
\end{document}